\documentclass[12pt]{amsart}
\usepackage {amssymb,amsmath}
\usepackage{graphicx}
\usepackage{hyperref}
\makeatletter
\def\@cite#1#2{{\m@th\upshape\bfseries%
[{#1\if@tempswa{\m@th\upshape\mdseries, #2}\fi}]}} \makeatother
\newtheorem{thm}{Theorem}[section]
\newtheorem{lem}[thm]{Lemma}
\newtheorem{cor}[thm]{Corollary}
\newtheorem{prop}[thm]{Proposition}
\newtheorem{rem}[thm]{Remark}

\newtheorem{defn}[thm]{Definition}

\newcommand{\Prf}{\noindent\textbf{Proof.\ }}
\newcommand{\bx}{\strut\hfill$\blacksquare$\medbreak}

\newcommand{\bbA}{{\mathbb{A}}}

\newcommand{\bbC}{{\mathbb{C}}}
\newcommand{\bbD}{{\mathbb{D}}}

\newcommand{\bbR}{{\mathbb{R}}}

 \newcommand{\C}{{\mathcal{C}}}
 
 \newcommand{\E}{{\mathcal{E}}}
 \newcommand{\F}{{\mathcal{F}}}
 
\renewcommand{\H}{{\mathcal{H}}}

\renewcommand{\L}{{\mathcal{L}}}

\renewcommand{\P}{{\mathcal{P}}}
 
 \newcommand{\R}{{\mathcal{R}}}

 \newcommand{\V}{{\mathcal{V}}}

\newcommand{\upchi}{{\raise.35ex\hbox{$\chi$}}}
\newcommand{\qiff}{\quad\text{if and only if}\quad}
\newcommand{\diag}{\operatorname{diag}}

\newcommand{\spn}{\operatorname{span}}

\newcommand{\conv}{\operatorname{conv}}
\newcommand{\spec}{\operatorname{spec}}
\def\bra#1{\langle #1 }
\def\ket#1{|#1 \rangle}
\def\kb#1#2{|#1\rangle\!\langle #2 |}

\begin{document}

\title[Higher-Rank Numerical Ranges]{Higher-Rank Numerical Ranges of Unitary and Normal Matrices}
\author[M.--D. Choi, J.A. Holbrook, D.W. Kribs, K. {\.Z}yczkowski]{Man-Duen
Choi$^1$, John A. Holbrook$^2$, David W. Kribs$^{2,3}$ \\ and
Karol {\.Z}yczkowski$^{4,5}$}

\thanks{{\it key words and phrases.} higher-rank numerical range, unitary matrix, quantum error
correction.}

\address{$^1$Department of Mathematics, University of Toronto,
Toronto, Ontario, Canada M5S 2E4}
\address{$^2$Department of Mathematics and Statistics,
University of Guelph, Guelph, Ontario, Canada  N1G 2W1}
\address{$^3$Institute for Quantum Computing, University of Waterloo, Waterloo,
ON, CANADA N2L 3G1}
\address{$^4$Institute of Physics, Jagiellonian University,
       ul. Reymonta 4, {30-059} Cracow, Poland}
\address{$^5$Center for Theoretical Physics, Polish Academy of Sciences,
   Al.~Lotnik{\'o}w 32/44, {02-668} Warsaw, Poland}

\begin{abstract}
We verify a conjecture on the structure of higher-rank numerical
ranges for a wide class of unitary and normal matrices. Using
analytic and geometric techniques, we show precisely how the
higher-rank numerical ranges for a generic unitary matrix are
given by complex polygons determined by the spectral structure of
the matrix. We discuss applications of the results to quantum
error correction, specifically to the problem of identification
and construction of codes for binary unitary noise models.
\end{abstract}
\maketitle

\section{Introduction}\label{S:introduction}

The study of higher-rank numerical ranges of matrices was
initiated in \cite{CKZ05a}, with a basic problem in quantum error
correction \cite{CKZ05b} giving the primary motivation.
Higher-rank numerical ranges generalize the classical numerical
range of a matrix, and arise as a special case of the matricial
range for a matrix \cite{Far93,LiTs91}. In \cite{CKZ05a}, three of
us conjectured that the higher-rank numerical ranges of normal
matrices depend, in a very precise way, on the spectral structure
of the matrix. The conjecture reduces in the rank-1 case to a
well-known property of the classical numerical range for a normal
matrix, and it has opened the door to some interesting new
mathematical problems. Its verification (or refutation) would also
yield information for quantum error correction.

In this paper we verify the higher-rank numerical range conjecture
for a wide variety of unitary and normal matrices. This is
accomplished by introducing a number of new geometric techniques
into the analysis. We show in the case of a generic $N\times N$
unitary with non-degenerate spectrum and positive integer $k\geq
1$ with $N\geq 3k$, that the $k$th numerical range is given by a
certain polygon in the complex plane determined by the eigenvalues
of the unitary, and thus verify the conjecture. In other cases,
such as $N=5m$ and $k=2m$, we also verify the conjecture. But the
analysis in these cases is more delicate, and our proof is
non-constructive in nature. Figure~1 below provides a chart
indicating the cases we verify, together with the various
constraints.

Our results may be applied to construct error correcting codes for
a special class of quantum channels. A ``binary unitary channel''
\cite{CKZ05b} $\E$ is a noise model described by two unitary
errors that can occur during the time evolution of a given quantum
system.  The construction of codes for such a channel relies on
the structure of the higher-rank numerical ranges for a single
unitary $U$. Suppose $U$ acts on $N$-dimensional Hilbert space.
Then an ``$[N,k]$-code'' for $\E$ is a $k$-dimensional subspace
code that is correctable for $\E$. The results described above for
$N\geq 3k$ yield a simple algorithm to determine the existence of
codes, and an explicit construction of codes when they exist.

The paper is organized as follows. Section 2 contains a discussion
of some basics of quantum error correction that give important
motivation for consideration of higher-rank numerical ranges. In
Section~3 we discuss the conjecture, and show in particular how
the general normal case relies on the unitary case. Section~4
deals with the structure of the aforementioned polygon and a
derivation of conditions under which it is nonempty. Section~5
includes the proof for $N\geq 3k$ and the construction of codes
for binary unitary channels. In Section~6 we derive a number of
other cases non-constructively based on the case $N=5$ and $k=2$.
We finish with a brief discussion on possible further avenues of
research and limitations in Section~7.

\section{Error Correction in Quantum Computing and Binary Unitary Channels}\label{S:ecruc}

For a more complete discussion on the material of this section see
\cite{CKZ05b} and the references therein. We start with a quantum
system in contact with an external environment having finitely
many degrees of freedom represented on a Hilbert space $\H$ such
that $\dim\H = N < \infty$. Consider a unitary time evolution of
the combined system and environment induced by a given Hamiltonian
associated with some quantum computation implemented on $\H$. The
action of the evolution map on the system is obtained by tracing
out the environment, and the resulting map is called a quantum
{\it channel} or {\it operation}. Such a map is described by a
completely positive, trace preserving map $\E:
\L(\H)\rightarrow\L(\H)$, and can always be represented in the
operator-sum form as $\E(\rho) = \sum_a E_a \rho E_a^\dagger$ for
a set of operators $\{E_a\}\subseteq\L(\H)$ satisfying $\sum_a
E_a^\dagger E_a = I$. As a convenience we shall write $\E =
\{E_a\}$ when the operators $E_a$ determine $\E$ in this way. The
$E_a$ are interpreted as the noise or errors induced by $\E$.

In the standard approach to quantum error correction, a {\it code}
on $\H$ is given by a subspace $\C\subseteq\H$ of dimension at
least two. Denote the projection of $\H$ onto $\C$ by $P_\C$. A
code $\C$ is (ideally) {\it correctable} for an operation $\E$ if
there is a quantum operation $\R:\L(\H)\rightarrow\L(\H)$ that
acts as a left inverse of $\E$ on $\C$;
\begin{equation}\label{leftinverse}
\big(\R\circ\E\big)(\rho) = \rho \quad\quad  \forall \rho \in P_\C
\,\L(\H) \, P_\C.
\end{equation}
Given a representation for $\E = \{E_a\}$, a code $\C$ is
correctable for $\E$ if and only if there are complex numbers
$(\lambda_{ab})$ such that
\begin{equation}\label{klcondition}
P_\C E_a^\dagger E_b P_\C = \lambda_{ab} P_\C \quad \forall \,
a,b.
\end{equation}
Thus, the problem of finding correctable codes for $\E$ is
equivalent to simultaneously solving the family of equations in
Eqs.~(\ref{klcondition}) for the scalars $\lambda_{ab}$ and
projections $P_\C$, for all pairs $a,b$.

Of central importance in quantum computing, communication and
cryptography is the class of {\it randomized unitary channels}
\cite{AL87,BZ06}. Such a channel has a representation of the form
$\E = \{ \sqrt{p_a} \, U_a\}$ where each $U_a$ is a unitary
operator and the $\{p_a\}$  form a classical probability
distribution; $p_a>0$, $\sum_a p_a =1$. Hence,
\begin{equation}\label{ruc}
\E(\rho) = \sum_a p_a U_a \rho U_a^\dagger \quad\quad \forall \rho
\in\L(\H).
\end{equation}
The associated quantum operation is given by the scenario in which
the error $U_a$ occurs with probability $p_a$. By
Eqs.~(\ref{klcondition}), finding ideal correctable codes for $\E
= \{ \sqrt{p_a} \, U_a\}$ is equivalent to solving the
(un-normalized) equations
\begin{equation}\label{unitarycondition}
P U_a^\dagger U_b P = \lambda_{ab} P \quad \forall \, a,b,
\end{equation}
for $\lambda_{ab}$ and $P$. Note that each operator $U_a^\dagger
U_b$ is unitary.

The following observation illustrates the importance of randomized
unitary channels in error correction.

\begin{prop}\label{rucprop}
Let $\E$ be a quantum operation. Then $\C$ is a correctable code
for $\E$ if and only if there is a randomized unitary channel $\F
$ such that $\C$ is correctable for $\F$ and
\begin{equation}\label{rucequal}
\E(\rho) = \F(\rho) \quad\quad \forall \rho\in P_\C \, \L(\H)
\,P_\C.
\end{equation}
\end{prop}

\Prf In fact, $\F= \{ \sqrt{p_a}\, U_a\}$ can be chosen so that
the partial isometries $U_a P_\C$ have mutually orthogonal ranges
for distinct $a$. This follows directly from the usual
construction of a correction operation $\R$ for $\E$ on $\C$
\cite{KL97a}. The matrix $(\lambda_{ab})$ from
Eqs.(\ref{klcondition}) is a density matrix, and the unitary which
diagonalizes it can be used to find a set of error operators
$\{F_a\}$ that implement $\E \circ \P_\C$, where
$\P_\C(\cdot)=P_\C(\cdot)P_\C$, such that $P_\C F_a^\dagger F_b
P_\C = \delta_{ab} d_{aa} P_\C$. The polar decomposition yields a
partial isometry $U_a$, which can be extended to a unitary on the
entire space, such that $F_a P_\C = U_a \sqrt{P_\C F_a^\dagger F_a
P_\C}= \sqrt{d_{aa}}\, U_a P_\C$. The $\{\sqrt{d_{aa}}\}$ form a
probability distribution by the trace preservation of $\E$.
 \bx

Of course, the randomized unitary channel given by the restriction
of $\E$ to the code $\C$ can only be obtained when the code itself
is known. Thus, this result in itself is of little practical use
for the general problem of finding error correcting codes.
Nevertheless, it indicates that the general problem is equivalent
to finding codes for the class of randomized unitary channels.

Observe that a code is correctable for a binary unitary noise
model $\E = \{ \sqrt{p}\, V, \sqrt{1-p}\,W\}$, where $V$ and $W$
are unitary, if and only if it is correctable for $\E = \{
\sqrt{p}\, I, \sqrt{1-p}\,V^\dagger W\}$.

\begin{defn}\label{bucdefn}
A {\bf binary unitary channel} on a Hilbert space $\H$ is a
channel of the form
\begin{equation}
\E = \{ \sqrt{p}\, I, \sqrt{1-p}\,U\}
\end{equation}
for some unitary $U\in\L(\H)$ and fixed probability $0<p<1$. Thus,
the action of $\E$ is given by
\begin{equation}\label{bucform}
\E(\rho) = p\, \rho + (1-p) \, U\rho U^\dagger \quad \quad \forall
\rho\in\L(\H).
\end{equation}
\end{defn}


\begin{rem}\label{bucremark}
Binary unitary channels form a rather restrictive class of
physical noise maps, but they provide a useful set of ``toy''
examples for testing the ``compression'' approach \cite{CKZ05b} to
build quantum error correcting codes enabled by consideration of
higher-rank numerical ranges. Observe that from
Eqs.~(\ref{klcondition}), the problem of finding ideal correctable
codes for a given binary unitary channel is equivalent to solving
four equations, but that
the entire problem reduces to solving the single (un-normalized)
equation for $\lambda$ and $P$ given by:
\begin{equation}\label{singleeqn}
PUP = \lambda\, P.
\end{equation}
An immediate consequence of what follows is an algorithm to
construct codes for a wide class of binary unitary channels. This
is encapsulated in the discussion of Section~5.
\end{rem}

\section{Higher-Rank Numerical Range Conjecture}\label{S:hrnrconj}

Given a fixed positive integer $k\geq 1$ and $T\in\L(\H)$, the
{\it $k$th numerical range} of $T$ is the set of complex numbers
\[
\quad\,\,\, \Lambda_k(T) = \big\{ \lambda\in\bbC : PT P = \lambda
P \,\, {\rm for \,\, some \,\, rank-}k \,\,{\rm projection\,\,} P
\big\}.
\]
The classical numerical range $W(T) = \Lambda_1(T)$ is obtained
when $k=1$.

\begin{defn}\label{omegadefn}
Let $T\in\L(\H)$ and let $k\geq 1$ be a fixed positive integer.
Then we define $\Omega_k(T)$ to be the intersection of the convex
hulls $\conv (\Gamma)$, where $\Gamma$ runs through all
$(N-k+1)$-point subsets (counting multiplicities) of the set of
eigenvalues $\spec(T)$ for $T$. That is,
\begin{eqnarray*}
\Omega_k(T) \,\,\,= \,\,\,\bigcap_{\Gamma\,\subseteq\,
\spec(T);\,\,|\Gamma|=N-k+1}\, \conv (\Gamma).
\end{eqnarray*}
\end{defn}

Thus, $\Omega_k(T)$ is a convex subset of the complex plane that
can be computed directly from the spectrum of $T$. Below we will
show how this set can typically be computed as an intersection of
much fewer than $N \choose N-k+1$ sets.

It is easy to see that $\Omega_k(T)$ contains $\Lambda_k(T)$ for
normal $T$. We include a short proof for completeness and
notational purposes.

\begin{prop}\label{easyinclusion}
Let $T\in\L(\H)$ be a normal operator and fix a positive integer
$k\geq 1$. Then $\Lambda_k(T)\subseteq\Omega_k(T)$.
\end{prop}

\Prf Let $\{\ket{\psi_1},\ldots,\ket{\psi_N}\}$ be a complete set
of orthonormal eigenvectors for $T$ with eigenvalues
$T\ket{\psi_j} = \lambda_j \ket{\psi_j}$. Let
$\lambda\in\Lambda_k(T)$ and let $P = \sum_{i=1}^k
\kb{\phi_i}{\phi_i}$ be a rank-$k$ projection such that $PT
P=\lambda P$. Then $\bra{T\phi_i}\ket{\phi_j} = \delta_{ij}
\lambda$ for all $i,j$. Let $\bbA$ be a subset of $\{1,\ldots
,N\}$ with cardinality $|\bbA| = k-1$. Choose a unit vector
$\ket{\phi}$ in the $k$-dimensional subspace $P\H =
\spn\{\ket{\phi_1},\ldots ,\ket{\phi_k}\}$ that is perpendicular
to all the $\ket{\psi_j}$ for which $j\in\bbA$; and so,
$\ket{\phi} = \sum_{j\notin\bbA} z_j \ket{\psi_j}$ with $\sum_j
|z_j|^2 =1$. Then we have
\begin{eqnarray}
\lambda = \bra{T\phi}\ket{\phi} = \sum_{j\notin\bbA} |z_j|^2
\lambda_j &\in& \conv\{\lambda_j :j\notin\bbA \},
\end{eqnarray}
and it follows that $\lambda$ belongs to $\Omega_k(T)$.
 \bx

It is well-known and easy to verify that the numerical range of a
normal operator $T$ coincides with the convex hull of its
eigenvalues (that is, $\Lambda_1(T)=\Omega_1(T)$). In
\cite{CKZ05a} the following conjecture was asserted as a
generalization of this fact.

{\noindent}{\bf Conjecture A.} Let $\H$ be an $N$-dimensional
Hilbert space and let $k\geq 1$ be a positive integer. Then for
every normal operator $T\in\L(\H)$,
\begin{equation}\label{conj}
\Lambda_k(T) \,\, = \,\, \Omega_k(T).
\end{equation}

The Hermitian case \cite{CKZ05a} and the normal $N\leq 4$ case
\cite{CKZ05b} of the conjecture have been verified previously. We
show that the general normal case of Conjecture~A can be reduced
to the unitary case.

\begin{prop}\label{reduction}
Conjecture~A holds if and only if the conjecture holds for all
unitary matrices.
\end{prop}

\Prf  First note that for a fixed $k$, a standard translation
argument shows the statement $\Lambda_k(T) = \Omega_k(T)$ for all
normal $T\in\L(\bbC^N)$ is equivalent to the statement
$0\in\Lambda_k(T)$ if and only if $0\in\Omega_k(T)$ for all normal
$T\in\L(\bbC^N)$. We focus on the latter formulation.

Every normal operator $T$ decomposes as $T = T_1 \oplus 0_m$,
where $T_1$ is normal and invertible. The case $m\geq k$ is easily
handled, so assume $m< k$. One can check that $ 0\in\Lambda_k(T)
\qiff 0\in\Lambda_{k-m}(T_1) $, and $ 0\in\Omega_k(T) \qiff
0\in\Omega_{k-m}(T_1). $ Let $\{\lambda_1,\ldots,\lambda_{N-m}\}$
be the (non-zero) eigenvalues for $T_1$, and let $U$ be the
unitary on $\bbC^{N-m}$ obtained from the polar decomposition of
$T_1$ with eigenvalues
$\{\frac{\lambda_1}{|\lambda_1|},\ldots,\frac{\lambda_{N-m}}{|\lambda_{N-m}|}\}$.
By assumption we have $\Lambda_{k-m}(U) = \Omega_{k-m}(U)$, and
hence zero belongs to both sets or neither set. Thus, we complete
the proof by showing that: (i) $0\in\Lambda_{k-m}(T_1) \qiff
0\in\Lambda_{k-m}(U)$, and (ii) $0\in\Omega_{k-m}(U) \qiff
0\in\Omega_{k-m}(T_1).$

Note that by invertibility we have $T_1 = UR=RU = \sqrt{R}\, U
\sqrt{R}$ where $R=\sqrt{T^\dagger T}\geq 0$ is invertible with
eigenvalues $\{|\lambda_1|,\ldots,|\lambda_{k-m}|\}$. Thus, (i)
follows from the more general principle that if $T=X^\dagger S X$
where $X$ is invertible, then $0\in\Lambda_k(T)$ if and only if
$0\in\Lambda_k(S)$. Indeed, if $P$ is a rank-$k$ projection such
that $PTP=0$, then the rank-$k$ range projection $Q$ of $XP$
satisfies $QSQ=0$.

For (ii), note that by definition $0\notin\Omega_{k-m}(T_1)$
precisely when 0 does not belong to the convex hull of $N-k+m+1$
of the eigenvalues $\{\lambda_1,\ldots,\lambda_{N-m}\}$. This is
equivalent to the existence of a line passing through the origin
that does not meet this convex hull. By the same argument, this
geometric condition is equivalent to $0\notin\Omega_{k-m}(U)$.
 \bx

This result, combined with the motivation from quantum computing
discussed above, naturally leads to a focus on the unitary case of
the conjecture. We introduce the following nomenclature to
delineate the generic unitary subcases.

\begin{defn}
We will use the notation ${\rm Conj}\,(N,k)$ to denote the
sub-conjecture of Conjecture~A given by the statement
Eq.~(\ref{conj}) for a given pair $[N,k]$ and all unitary
operators on $N$-dimensional Hilbert space with non-degenerate
spectrum. Further, we will say ${\rm Conj}\,(N,k)$ is {\it
constructively verified} for a given pair $[N,k]$ if it is shown
that Eq.~(\ref{conj}) holds for every unitary operator $U$ on $\H
= \bbC^N$, and if, whenever $\Lambda_k(U)$ is nonempty, for every
$\lambda\in\Lambda_k(U)$ a rank-$k$ projection $P$ can be
explicitly constructed such that $P U P = \lambda P$.
\end{defn}


\begin{figure} [htbp]
       \begin{center} \
     \includegraphics[width=10cm,angle=0]{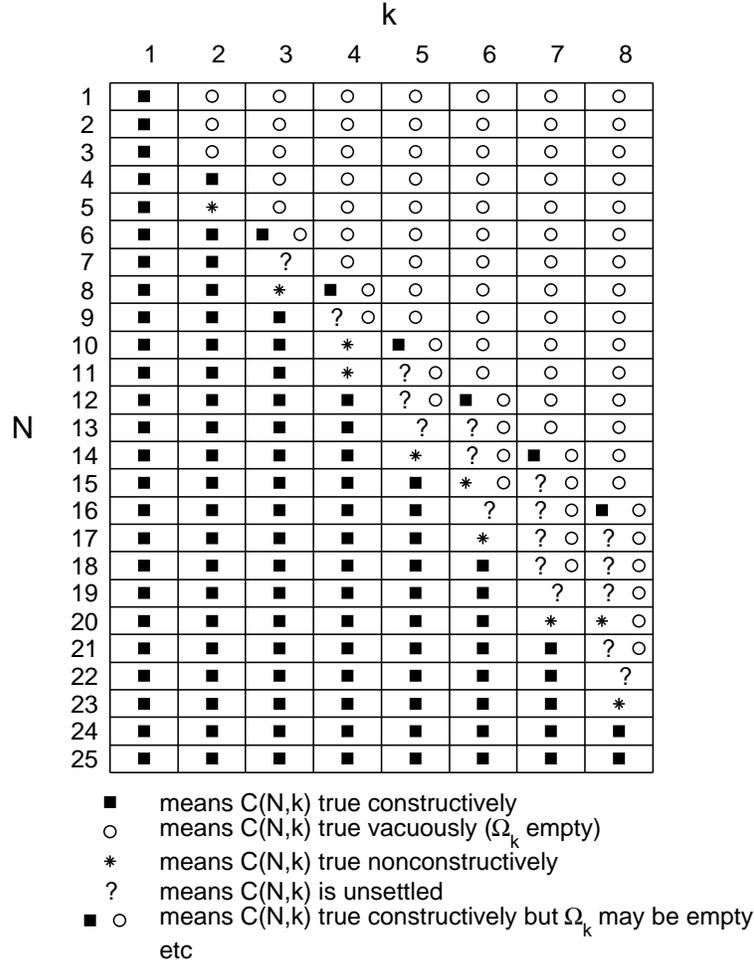}
\caption{Conj($N,k$) for nondegenerate unitary $U$.} \label{fig1}
\end{center}
     \end{figure}

\section{The Structure of $\Omega_k$}\label{S:omegastructure}

In this section we analyse the geometric structure of the set
$\Omega_k(U)$ for a generic unitary $U$ on $N$-dimensional Hilbert
space with non-degenerate spectrum. Let us first establish
notation we will use for the rest of the paper.

We shall consider the case of a unitary $U$ with eigenvalues
$\lambda_j = \exp(i \theta_j)$, $j=1,\ldots ,N$, such that $0 \leq
\theta_1 < \theta_2 < \ldots < \theta_N < 2 \pi$. Thus, the
eigenvalues $\lambda_j$ are ordered counterclockwise around the
unit circle $\partial\bbD$ in $\bbC$ (we use $\bbD$ to denote the
closed unit disc). For multiple eigenvalues the numbering is
arbitrary, but we choose an orthonormal system of eigenvectors
$\ket{\psi_j}\in\H$ such that
\begin{equation}
 U\ket{\psi_j} = \lambda_j\ket{\psi_j}.
\end{equation}
When appropriate we extend the numbering of the $\lambda_j$ and
$\ket{\psi_j}$ cyclically: for example, $\lambda_{N+1}$ means
$\lambda_1$. Given integers $i,j$ with $i<j\leq i+N$,
let $D(i,j,U)$ denote the compact convex subset of $\bbC$ bounded
by the line segment from $\lambda_i$ to $\lambda_j$ and the
counterclockwise circular arc from $\lambda_j$ to $\lambda_i$;
recall our conventions about cyclical numbering of the
$\lambda_j$. We interpret $D(i,i+N,U)$ as $\{\lambda_i\}$.

We first show that the form of $\Omega_k(U)$ is simpler than what
Definition~\ref{omegadefn} suggests. In particular, $\Omega_k(U)$
is a filled convex polygon with at most $N$ sides.

\begin{lem}\label{omegaform}
For all $k\geq 1$ and every unitary $U\in\L(\bbC^N)$, the set
$\Omega_k(U)$ is the convex polygon given by
\begin{equation}\label{omegaformeqn}
\Omega_k(U)\,=\,\bigcap_{i=1}^N D(i,i+k,U).
\end{equation}
\end{lem}

\Prf Let $\Omega$ denote the intersection of
Eq.~(\ref{omegaformeqn}). The cardinality of the set
\[
   S=\{1,2,\dots,N\}\setminus\{i+1,i+2,\dots,i+k-1\}
\]
(where integers are interpreted modulo $N$) is $|S|=N-k+1$. Thus
\[
   \Omega_k(U)\subseteq\conv(\{\lambda_j:j\in S\})\subseteq D(i,i+k,U).
\]
Since this holds for all $i$, we have
$\Omega_k(U)\subseteq\Omega$.

On the other hand, if $|S|=N-k+1=s$ we may write
$S=\{i_1,i_2,\dots,i_s\}$ with $1\leq i_1<i_2<\dots<i_s\leq N$ and
\[
   \conv(\{\lambda_j:j\in S\})=\bigcap_{j=1}^s D(i_j,i_{j+1},U)
\]
(it is understood here that $i_{s+1}=i_1$). Since between $i_j$
and $i_{j+1}$ there are $i_{j+1}-i_j-1$ integers that are omitted
from $S$, we must have $i_{j+1}-i_j\leq k$, so that
$D(i_j,i_{j+1},U)\supseteq D(i_j,i_j+k,U)$. It follows that
$\Omega$ is contained in $\conv(\{\lambda_j:j\in S\})$ for each
such $S$. Hence $\Omega_k(U)\supseteq\Omega$, and equality is
verified.
 \bx

The following containments are direct consequences of this result.

\begin{cor}\label{omegacontain1}
For all $U$ and all $k$, we have
$\Omega_{k+1}(U)\subseteq\Omega_k(U)$.
\end{cor}

\Prf In view of Lemma~\ref{omegaform}, we need only observe that,
for any $i$, $D(i,i+k+1,U)\subseteq D(i,i+k,U)$.
 \bx

\begin{cor}\label{omegacontain2}
If $V$ is unitary and $\spec(U)\subseteq\spec(V)$, then
$\Omega_k(U)\subseteq\Omega_k(V)$ for all $k$.
\end{cor}

\Prf It is enough to treat the case where
$\spec(V)=\spec(U)\cup\{\lambda_{N+1}\}$ and to arrange the
geometric ordering so that
$\spec(U)=\{\lambda_1,\lambda_2,\dots,\lambda_N\}$ and
$\spec(V)=\{\lambda_1,\lambda_2,\dots,\lambda_{N+1}\}$. For $1\leq
i\leq N-k$ we have $D(i,i+k,V)=D(i,i+k,U)$; for $N-k<i\leq N$, we
have $D(i,i+k,V)\supseteq D(i,i+k,U)$ (note that the extended
numbering within $\spec(V)$ is done modulo $N+1$); finally,
$D(N+1,N+1+k,V)\supseteq D(N,N+k,U)$. Using Lemma~\ref{omegaform},
we see that $\Omega_k(V)\supseteq\Omega_k(U)$.
 \bx

The following corollary points out that Conj($N,k$) is easy to
verify when $k$ divides $N$.

\begin{cor}\label{kdividesN}
Suppose that $k$ divides $N$, with $m=N/k$. Then $\Omega_k(U)$ is
the intersection of $k$ $m$--gons, and Conj($N,k$) follows.
\end{cor}

\begin{figure} [htbp]
       \begin{center} \
     \includegraphics[width=8cm,angle=0]{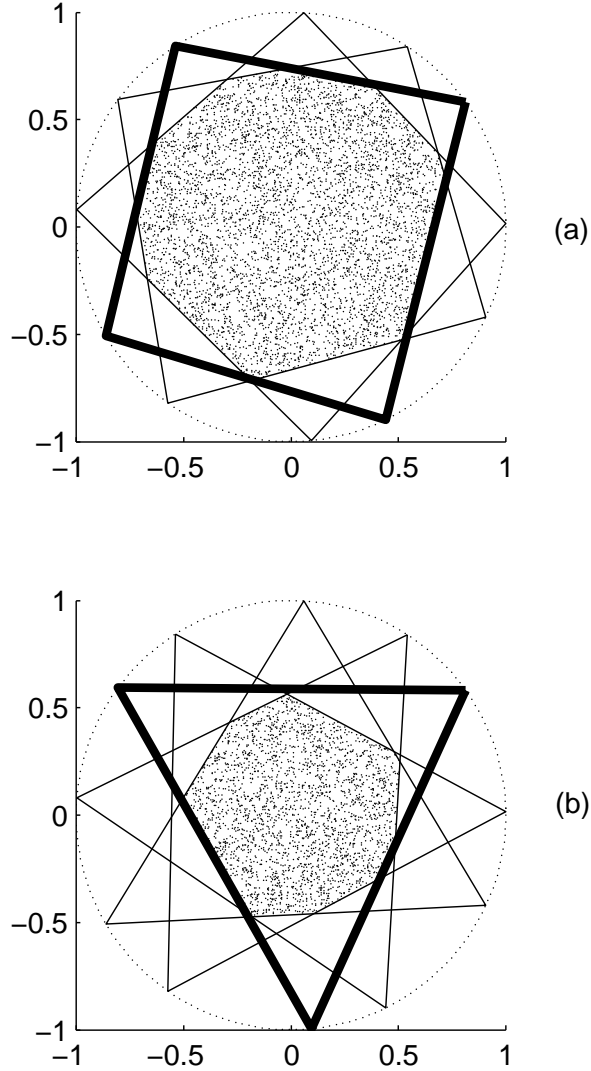}
\caption{Corollary 4.4 in action: (a) $\Omega_3(U)$ as
intersection of 3 quadrilaterals when $N=12$; (b) $\Omega_4(U)$ as
intersection of 4 triangles when $N=12$.} \label{fig2}
\end{center}
     \end{figure}

\Prf Let $S_i=\{i,i+k,i+2k,\dots,i+(m-1)k\}$ ($i=1,2,\dots,k$);
these index sets partition $\{1,2,\dots,N\}$. In view of Lemma
4.1, the $m$--gons
\[ G_i=\bigcap_{j\in S_i}D(i,i+k,U) \]
intersect to form $\Omega_k(U)$. Now consider
$\lambda\in\Omega_k(U)$. Since $\lambda\in G_i$ for each $i$ we
may write
\[ \lambda=\sum_{j\in S_i} t_{ij}\lambda_j \]
as a convex combination ($t_{ij}\geq0$, $\sum_{j\in
S_i}t_{ij}=1$). For $i=1,2,\ldots ,k$, let
\[
    \ket{\phi_i}=\sum_{j\in S_i}\sqrt{t_{ij}}\ket{\psi_j};
\]
clearly the $\ket{\phi_i}$ are unit vectors, and they are
orthogonal because the $S_i$ are disjoint. We see that $(U-\lambda
I)\ket{\phi_i}\perp\ket{\phi_{i'}}$ for all $i,i'$, so that
$\V=\spn\{\ket{\phi_i}:i=1,2,\dots,k\}$ satisfies $(U-\lambda
I)\V\perp \V$. We can thus define the rank-$k$ projection $P =
\sum_{i=1}^k \kb{\phi_i}{\phi_i}$ onto $\V$. Then $P(U-\lambda
I)P=0$ and it follows that $\lambda\in\Lambda_k(U)$. Together with
Proposition~\ref{easyinclusion}, we have verified Conj($N,k$) in
these cases. \bx


\begin{rem} The corollary above is usually of interest when $m\geq3$. If
$N=k$ ($m=1$), then $\Omega_k(U)$ will be empty unless $U$ is a
scalar. If $N=2k$ ($m=2$), then $D(i,i+k,U)\cap D(i+k,i+N,U)$ is
the line segment $[\lambda_i,\lambda_{i+k}]$. Thus $\Omega_k(U)$
is usually empty in this case for $k>2$; in cases where $\spec(U)$
has a special symmetry, $\Omega_k(U)$ may be a single point.
\end{rem}

Under certain conditions Corollary 4.3 may be made more precise.

\begin{prop}\label{refinedomegaform}
Suppose $V$ is unitary on $N+1$-dimensional Hilbert space and
$\spec(V)=\{\lambda_1,\lambda_2,\dots,\lambda_{N+1}\}$ with
distinct $\lambda_j$. If $\Omega_{k+1}(V)\neq\emptyset$, then
\begin{equation}
    \Omega_k(V)=\bigcup_{j=1}^{N+1}
 \,\,   \Omega_k(V_j),
\end{equation}
where $V_j$ is the unitary with spectrum
$\spec(V)\setminus\{\lambda_j\}$.
\end{prop}

\Prf Arguing as in the proof of Corollary 4.3 we see that
$\Omega_k(V_{N+1})=\Omega_k(V)\cap C(N+1)$, where
\[
   C(N+1)=\bigcap_{j=N+1-k}^N D(j,j+k+1,V).
\]
Let $\lambda$ be a point in $\Omega_{k+1}(V)$ (which is nonempty
by hypothesis). Then $\lambda\in C(N+1)$ by Lemma~\ref{omegaform}.
Note also that $C(N+1)$ includes  the counterclockwise arc
$A(N+1)$ of $\partial\bbD$ from $\lambda_{N+k+1}=\lambda_k$ to
$\lambda_{N+1-k}$; since $N+1-k>k$ (otherwise we would have
$N+1\leq 2k<2(k+1)$ so that $\Omega_{k+1}(V)$ would be empty), the
arc $A(N+1)$ has nonempty interior (relative to $\partial\bbD$).
Likewise, for each $j=1,2,\dots,N+1$ we have
$\Omega_k(V_j)=\Omega_k(V)\cap C(j)$ where $C(j)$ is a convex set
containing $\lambda$ and an arc $A(j)$. It remains to show that
$\bigcup_j C(j)=\bbD$. The arcs $A(j)$ (overlapping in general)
cover all of $\partial\bbD$ and each $C(j)$ includes the
``sector'' $\conv(\{\lambda\}\cup A(j))$; these sectors certainly
cover $\bbD$.
 \bx

The  previous result raises the general question of which
$\Omega_k(U)$ are nonempty. This question is of course important
for the identification and construction of error-correcting codes.
In particular, Conj($N,k$) may include cases where both
$\Lambda_k(U)$ and $\Omega_k(U)$ are empty, but this is not
helpful for the applications. The question is somewhat clarified
by the following.

\begin{thm}\label{omegastructure}
Let $U\in\L(\bbC^N)$ be a unitary with $N$ distinct eigenvalues
and let $k\geq 1$. Then we have the following conditions on
$\Omega_k(U)$.

(1) If $N< 2k$, then $\Omega_k(U)=\emptyset$.

(2) If $N= 2k$, then $\Omega_k(U)$ is empty if the line segments
$[\lambda_j,\lambda_{j+k}]$ do not intersect, and, otherwise, it
is the singleton set given by the intersection point of these line
segments.

(3) If $2k < N < 3k-2$, then $\Omega_k(U)$ can be either empty or
non-empty.

For any unitary $U\in\L(\bbC^N)$, whether or not the eigenvalues
are distinct, we have:

(4) If $N\geq 3k-2$, then $\Omega_k(U)$ is always nonempty.
\end{thm}

\begin{figure} [htbp]
       \begin{center} \
     \includegraphics[width=8cm,angle=0]{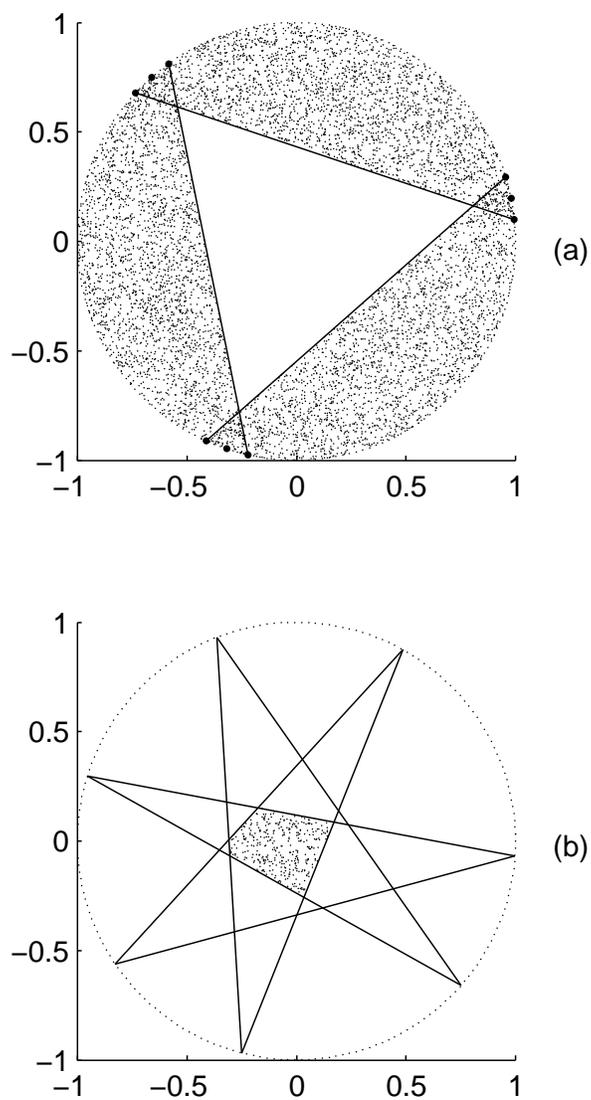}
\caption{Illustrating Theorem 4.7: (a) Here $N=9$ and $k=4$; since
$N<3k-2$ it may happen, as here, that $\Omega_4(U)=\emptyset$;
note that $D(3,7,U)$, $D(6,1,U)$, and $D(9,4,U)$ overlap pairwise
but have no common point. (b) Here $N=7$ and $k=3$; since $N\geq
3k-2$, the core $\Omega_3(U)$ of this 7--pointed star must be
nonempty.} \label{fig3}
\end{center}
     \end{figure}

\Prf  To see (1) note that $D(j,j+k,U)\cap D(j+k,j+2k,U)$ is the
singleton $\{\lambda_{j+k}\}$ and that the $\{\lambda_{j+k}\}$ are
distinct. See also Proposition~1 from \cite{CKZ05b}. The case (2)
is a special case of Corollary~\ref{kdividesN}. Here,
$\Omega_k(U)$ is non-empty (and a singleton set) precisely when
the line segments $[\lambda_j,\lambda_{j+k}]$ intersect in a
common point.

In view of Corollary~\ref{omegacontain2}, it is sufficient for the
statement (3) to provide an example of $U$ with $N=3(k-1)$ and
$\Omega_k(U)=\emptyset$. Such an example is provided by grouping
$k-1$ of the $\lambda_j$ close to and on the counterclockwise side
of each of the cube roots of unity. It is then clear that
\[
   D(k-1,2k-1,U)\cap D(2k-2,3k-2,U)\cap D(3k-3,4k-3,U)=\emptyset;
\]
note that here $N=3k-3$ so that $3k-2\equiv 1$ and $4k-3\equiv k$.
Again recalling Lemma~\ref{omegaform}, we see that \emph{a
fortiori} $\Omega_k(U)=\emptyset$.

Concerning the statement (4), we recall Helly's Theorem from
convex analysis (see, for example, Chapter 3 of \cite{Lay}): a
family of compact, convex subsets of $\bbR^d$ has nonempty
intersection provided any subfamily of size $d+1$ has nonempty
intersection. Since each $D(i,i+k,U)$ is compact and convex in
$\bbR^2\equiv\bbC$ we need only prove that $N \geq 3k-2$ implies
\[
   D(a,a+k,U)\cap D(b,b+k,U)\cap D(c,c+k,U)\neq\emptyset,
\]
then invoke Helly's Theorem for $d=2$ and Lemma~\ref{omegaform}.
Such a triple intersection could only be empty if the complements
in $\bbD$ covered all of $\bbD$. In particular, the omitted arcs
(strictly between $\lambda_a$ and $\lambda_{a+k}$, etc) would
cover $\partial\bbD$. Each of these arcs contains $k-1$ points
from $\spec(U)$, so we would have the contradiction $N\leq
3(k-1)$. \bx


\section{Verification of the Conjecture for $N\geq 3k$ and Construction of Codes for Binary Unitary Channels}\label{S:constructiveverify}

In this section we give a constructive verification of Conj($N,k$)
in the cases $N\geq 3k$. In such cases we explicitly construct
error-correcting codes. We require more notation.

Let $\Delta_k(U)$ denote the set of those $\lambda\in\bbC$ such
that for some $k$ \emph{disjoint} subsets $S_1,S_2,\dots,S_k$ of
$\{1,2,\dots,N\}$
we have $\lambda\in\conv(\{\lambda_j:j\in S_i\})$ for each $i$.
The $\Delta$ in the notation $\Delta_k(U)$ is to recall the
``disjoint'' condition. Evidently
$\Delta_k(U)\subseteq\Delta_k(V)$ whenever
$\spec(U)\subseteq\spec(V)$.

A straightforward generalization of the construction used in
Corollary 4.4 yields the following result.

\begin{lem}\label{deltalambda}
For every unitary $U\in\L(\bbC^N)$ and $k\geq 1$ we have
$\Delta_k(U)\subseteq\Lambda_k(U)$.
\end{lem}

\Prf Let $\lambda\in\Delta_k(U)$ be expressed as a convex
combination of the eigenvalues for $U$ for each $i$:
\[
   \lambda=\sum_{j\in S_i} t_{ij}\lambda_j,\quad t_{ij}\geq
   0,\quad \sum_{j\in S_i} t_{ij}=1.
\]
Let $\ket{\phi_i}=\sum_{j\in S_i}\sqrt{t_{ij}}\ket{\psi_j}$. Then
$\ket{\phi_1},\dots,\ket{\phi_k}$ are orthonormal (orthogonal
because the $S_i$ are disjoint). Let $P$ be the orthogonal
rank-$k$ projection onto
$\V=\spn\{\ket{\phi_1},\dots,\ket{\phi_k}\}$. Then for each
$\ket{\psi}\in \V$, the vector $(U-\lambda I)\ket{\psi}$ is
orthogonal to $\V$. Indeed, if $i\neq i'$ then $(U-\lambda
I)\ket{\phi_i}\in\spn\{\ket{\psi_j}:j\in S_i\}$, which in turn is
orthogonal to $\ket{\phi_{i'}}$ because
$\ket{\phi_{i'}}\in\spn\{\ket{\psi_j}:j\in S_{i'}\}$ and $S_i\cap
S_{i'}=\emptyset$.  On the other hand, $\bra{(U-\lambda
I)\phi_i}\ket{\phi_i} =\bra{U\phi_i}\ket{\phi_i} -\lambda$ and
\[
   \bra{U\phi_i}\ket{\phi_i}=\bra{\sum_{j\in S_i} \lambda_j\sqrt{t_{ij}}\psi_j}
   \ket{\sum_{j\in S_i}\sqrt{t_{ij}}\psi_j}=\sum_{j\in S_i} t_{ij}\lambda_j=\lambda.
\]
Hence $PUP=\lambda P$; that is, $\lambda$ belongs to
$\Lambda_k(U)$.
 \bx

The converse inclusion holds in a wide variety of cases.

\begin{thm}\label{N3k}
If $N\geq 3k$, then $\Omega_k(U)=\Delta_k(U)$. Hence, Conj($N,k$)
holds whenever $N\geq 3k$.
\end{thm}

\Prf Let $T(a,b,c)$ denote the ``eigentriangle''
$\conv(\{\lambda_a,\lambda_b,\lambda_c\})$. We shall say
$T(a,b,c)$ and $T(a',b',c')$ are ``disjoint'' if the index sets
$\{a,b,c\}$ and $\{a',b',c'\}$ are disjoint; of course, the
eigentriangles themselves may very well overlap. We show that
$N\geq3k$ implies that every $\lambda\in\Omega_k(U)$ lies in $k$
(pairwise) disjoint eigentriangles, and so
$\lambda\in\Delta_k(U)\subseteq\Lambda_k(U)$ by
Lemma~\ref{deltalambda}. This is clear for $k=1$ since any convex
polygon (here $\Omega_1(U)$, that is
$\conv(\{\lambda_j:j=1,2,\dots,N\}$) is (in many ways) a union of
triangles formed from the vertices.

For $k>1$ we proceed by induction. The wedge $W=D(1,k+1,U)\cap
D(N-k+1,1,U))$ contains $\Omega_k(U)$. Consider the eigentriangles
\begin{eqnarray*}
T(1,k+1,2k+1),\quad \dots \quad ,T(1,k+(N-3k+1),2k+(N-3k+1));
\end{eqnarray*}
note that (because $N\geq3k$) $2k+1\leq N-k+1$ and that the last
eigentriangle in this list is $T(1,N-2k+1,N-k+1)$. Thus the union
of these overlapping eigentriangles covers the part of $W$ that
contains $\Omega_k(U)$, and hence covers $\Omega_k(U)$ itself.
Given any $\lambda\in\Omega_k(U)$, choose one of the
eigentriangles from this list that contains $\lambda$. Let the
chosen eigentriangle be $T(1,b,c)$; note that $b\geq k+1$,
$c=b+k$, and $c\leq N-k+1$. We claim that $\lambda$ is also in
$\Omega_{k-1}(W)$, where $\spec(W)=\{\lambda_j:j\neq1,b,c\}$.
Assuming this claim is correct for the moment, we see that the
inductive step is achieved, since $|\spec(W)|=N-3\geq3(k-1)$ so
that $\lambda$ lies in $k-1$ disjoint eigentriangles drawn from
$\spec(W)$ as well as in $T(1,b,c)$, and hence that
$\lambda\in\Delta_k(U)$.

To verify the claim keep Lemma~\ref{omegaform} in mind and note
that the sets $D(j,j+k-1,W)$ that intersect to form
$\Omega_{k-1}(W)$ strictly contain one of the $D(i,i+k,U)$ unless
the arc omitted from $D(j,j+k-1,W)$ includes one of
$\lambda_1,\lambda_b,\lambda_c$, in which case $D(j,j+k-1,W)$
coincides with one of the $D(i,i+k,U)$. The key point is that the
arc cannot contain \emph{more} than one of
$\lambda_1,\lambda_b,\lambda_c$ since these are separated by at
least $k-1$ points in $\spec(W)$. Thus, in fact,
$\Omega_{k-1}(W)\supseteq\Omega_k(U)$, and this completes the
proof. \bx

Let us discuss the construction of codes. The cases in which $k$
divides $N$ are perhaps the simplest cases in which codes can be
explicitly constructed, as described in the proof of
Corollary~\ref{kdividesN}. The proof of Conj($N,k$) for $N\geq 3k$
given in Theorem~\ref{N3k} is also constructive in the sense that
the $\lambda\in\Lambda_k(U)$ and the corresponding projections $P$
may be found explicitly by an algorithm based on the proof. We
state this in terms of the general binary unitary channel error
correction problem.

Let $U$ be a unitary on $\bbC^N$ and let $k$ be a positive integer
such that $N\geq 3k$. Then by Theorem~\ref{omegastructure}~(4) and
Theorem~\ref{N3k}, we have $\Delta_k(U) = \Lambda_k(U) =
\Omega_k(U)$ and this set is nonempty. Following the proof of
Theorem~\ref{N3k} (and recalling our earlier notation), a
$k$-dimensional correctable code for any channel of the form $\E =
\{ \sqrt{p}\,I, \sqrt{1-p}\,U\}$ can be constructed by:

$(i)$ Compute $\Omega_k(U)$ from the eigenvalues
$\{\lambda_1,\ldots,\lambda_N\}$ of $U$. This can be done by using
Lemma~\ref{omegaform} in general, or by simpler means in special
cases, such as that of Corollary~\ref{kdividesN}.

$(ii)$ Choose $\lambda\in\Omega_k(U)$. By Theorem~\ref{N3k}, we
can find $k$-eigentriangles $T(a_j,b_j,c_j)$, $j=1,\ldots,k$, such
that each contains $\lambda$ and there are no repeats in the set
$\{a_j,b_j,c_j\}_{j=1}^k \subseteq \{1,\ldots,N\}$.

$(iii)$ As $\lambda$ belongs to each of the convex hulls
$\conv\{\lambda_{a_j},\lambda_{b_j},\lambda_{c_j}\}$, for
$j=1,\ldots,k$, we can compute $t_{1j}, t_{2j}, t_{3j} \geq 0$,
$\sum_{i=1}^3 t_{ij} =1$ such that $\lambda = t_{1j} \lambda_{a_j}
+ t_{2j} \lambda_{b_j} + t_{3j} \lambda_{c_j}$.

$(iv)$ For $j=1,\ldots,k$, put $s_{ij}=\sqrt{t_{ij}}$ and define
(orthonormal) states $\ket{\phi_j} = s_{1j} \ket{\psi_{a_j}} +
s_{2j} \ket{\psi_{b_j}} + s_{3j} \ket{\psi_{c_j}}$. Let $P =
\sum_{j=1}^k \kb{\phi_j}{\phi_j}$. Then $P\bbC^N =
\spn\{\ket{\phi_1},\ldots,\ket{\phi_N}\}$ is a $k$-dimensional
correctable code for $\E$.

\begin{rem}
It is instructive to rephrase this construction in terms of the
number of qubits, assuming that the entire system has dimension
$N=2^n$. The codes constructed above work for $k = [N/3]$, which
for $n\geq 2$ is not smaller than $N/4=2^{n-2}$. Therefore these
error-correcting codes support $n-2$ logical qubits. Our
construction shows that such codes are parametrized by the complex
numbers $\lambda\in\Lambda_k(U)$, which is a nonempty set
explicitly determined by the eigenvalues for $U$ as we have shown.
This result is optimal for $n\geq 3$ in the sense that for a
generic unitary one cannot obtain a code preserving $n-1$ qubits.
This observation follows from the fact that $\Lambda_{N/2}(U)$ is
not empty only in a very specific situation; if the $N/2$ lines
joining opposite (with respect to the ordering number of the
phase) eigenvalues of $U$ cross in a single point.

This result considered in the context of qudits (which denote
$d$-level systems) has the following implication: If $N=d^n$ and
$d\geq 3$ then $k=[N/3]\geq N/d$, so the constructed code supports
$d-1$ qudits.
\end{rem}


\section{Non-Constructive Verifications of the Conjecture}\label{S:nonconstructiveverify}

In this section we derive a non-constructive verification of
Conj($N,k$) in the case $N=5$, $k=2$, and then we extend the proof
to a variety of cases. For convenience we consider only $U$ with
distinct eigenvalues.

To move beyond the limitations of $\Delta_k(U)$ we introduce
$\Sigma_k(U)$ as the set of all $\lambda\in\conv(\spec(U))$ such
that for some single convex combination
\[
   \lambda=\sum_j t_j \lambda_j,\quad t_j\geq0,\quad \sum_j t_j=1
\]
we have $\alpha_{ij}\in\bbC$ ($i=1,2,\dots,k$; $j=1,2,\dots,N$)
with the following properties: $|\alpha_{ij}|=\sqrt{t_j}$ for all
$i$ and $j$, and whenever $i\neq i'$ we have both
$\sum_j\alpha_{ij}\overline{\alpha_{i'j}}=0$ and
$\sum_j\lambda_j\alpha_{ij}\overline{\alpha_{i'j}}=0$. The symbol
$\Sigma$ in $\Sigma_k(U)$ is chosen to recall that we use a
\emph{single} convex combination to represent each $\lambda$.

It is easy to see that $\Sigma_k(U)\subseteq\Lambda_k(U)$:
consider any $\lambda\in\Sigma_k(U)$ and (using again the above
notation) let $\ket{\phi_i}=\sum_j\alpha_{ij}\ket{\psi_j}$. The
conditions on the $\alpha_{ij}$ directly imply that the
$\ket{\phi_i}$ are orthonormal, that
$\bra{U\phi_i}\ket{\phi_i}=\lambda$ for each $i$, and that
$\bra{U\phi_i}\ket{\phi_{i'}}=0$ whenever $i\neq i'$. Let $P$ be
the rank-$k$ orthogonal projection onto the subspace
$\V=\spn\{\ket{\phi_1},\dots,\ket{\phi_k}\}$. Clearly $(U-\lambda
I)\V$ is orthogonal to $\V$, so that $PUP=\lambda P$ and
$\lambda\in\Lambda_k(U)$.

Let $\sigma$ denote $\spec(U)$ considered as a vector
$(\lambda_1,\lambda_2,\dots,\lambda_N)\in\bbC^N$. In terms of
$\sigma$ the following lemma provides a recipe for making elements
of $\Sigma_2(U)$.

\begin{lem}\label{sigmalemma}
Given $p\in\bbC^N$ such that $\vec0\neq
p\perp\{\vec1,\sigma,\overline{\sigma}\}$, set $s(p)=\sum_j|p_j|$
and
\[
   f(p)=\sum_j\frac{|p_j|}{s(p)}\lambda_j.
\]
Then $f(p)\in\Sigma_2(U)$.
\end{lem}

\Prf Let $\alpha_{1j}=\sqrt{|p_j|/s(p)}$ and
$\alpha_{2j}=\alpha_{1j}\overline{p_j}/|p_j|$ (if $p_j=0$, let
$\alpha_{2j}=\alpha_{1j}$ ($=0$)). With $\lambda=f(p)$ and
$t_j=|p_j|/s(p)$ we easily verify the requirements for the
$\alpha_{ij}$. For example, $\sum_j
\lambda_j\alpha_{1j}\overline{\alpha_{2j}}=(\sum_j
\lambda_jp_j)/s(p)=(p,\overline{\sigma})/s(p)=0$.
 \bx

\begin{cor}\label{sigma2cor}
For any $k\geq2$ we have $\Delta_k(U)\subseteq\Sigma_2(U)$.
\end{cor}

\Prf Consider $\lambda\in\Delta_k(U)$. Using again the notation
from the proof of Lemma 5.1, let
$t_{1*}=(t_{11},t_{12},\dots,t_{1N})$ with the understanding that
$t_{1j}=0$ if $j\not\in S_1$; similarly define $t_{2*}$. Let
$p=t_{1*}-t_{2*}$. Then $p\neq\vec0$ since the supports of
$t_{1*}$ and $t_{2*}$ are disjoint;
$(p,\vec1)=\sum_jt_{1j}-\sum_jt_{2j}=1-1=0$;
$(p,\sigma)=\sum_jt_{1j}\overline{\lambda_j}-
\sum_jt_{2j}\overline{\lambda_j}=
\overline{\lambda}-\overline{\lambda}=0$; similarly $p$ is
perpendicular to $\overline{\sigma}$. Thus $f(p)\in\Sigma_2(U)$.
Finally, since the supports of $t_{1*}$ and $t_{2*}$ are disjoint,
$f(p)=(\sum_jt_{1j}\lambda_j+\sum_jt_{2j}\lambda_j)/2=(\lambda+\lambda)/2=\lambda$.
 \bx

\begin{rem} Although Corollary 6.2 is sufficient for our present needs,
it should be noted that in fact $\Delta_k(U)\subseteq\Sigma_k(U)$
for any $k$. Indeed, for $\lambda\in\Delta_k(U)$ there are
$t_{ij}\geq0$ ($i=1,2,\dots,k;\,\,j=1,2,\dots,N$) such that
$\sum_j t_{ij}=1$, $\sum_j t_{ij}\lambda_j=\lambda$, and for each
$j$ at most one of the $t_{ij}$ is nonzero. Thus, setting
$s_{ij}=\sqrt{t_{ij}}$ we have a $k\times N$ matrix $S$ such that
$S\diag(\lambda_j)S^*=\lambda I_k$ and $SS^*=I_k$. Let $F$ be any
$k\times k$ unitary matrix with $|f_{ij}|=1/\sqrt{k}$ (for all
$i,j$); for example, $F$ could be the ``finite Fourier
transform'', where $f_{ij}=\omega^{ij}/\sqrt{k}$ with $\omega$ a
primitive $k$--th root of unity. Setting $R=FS$ we have
$R\diag(\lambda_j)R^*=F\lambda I_kF^*=\lambda I_k$, and
$RR^*=I_k$. Thus we may verify that $\lambda\in\Sigma_k(U)$ by
considering $\alpha_{ij}=r_{ij}$, since (for each $j$)
$|\alpha_{ij}|$ is then independent of $i$; in fact, for the given
$j$ just one $s_{mj}\neq0$ so that
$|\alpha_{ij}|=|f_{im}s_{mj}|=|s_{mj}|/\sqrt{k}$.
\end{rem}

It is clear that when $N=5$ only the boundary
$\partial\Omega_2(U)$ is captured by $\Delta_2(U)$; each point on
an edge of the pentagon $\Omega_2(U)$ belongs both to a line
segment $\conv(\{\lambda_i,\lambda_{i+2}\})$ and to the
complementary eigentriangle $T(i-1,i+1,i+3)$. To capture the
interior of $\Omega_2(U)$ we turn to $\Sigma_2(U)$. The
topological techniques used in the proof of the following result
are interesting mathematically but make the construction of
specific projections $P$ corresponding to $\lambda\in\Lambda_2(U)$
more difficult than we saw with techniques based on $\Delta_k(U)$.

\begin{thm}\label{52thm}
If $N=5$, then $\Omega_2(U) =\Sigma_2(U)$. Thus, Conj(5,2) is
correct.
\end{thm}

\Prf By Corollary~\ref{sigma2cor} we know that
$\partial\Omega_2(U)=\Delta_2(U)\subseteq\Sigma_2(U)$; to capture
the interior we first elaborate the ideas in the proof of that
corollary. Let $a_i$ be the vertex of $\Omega_2(U)$ at the point
of intersection between the line segment
$[\lambda_i,\lambda_{i+2}]$ (in other words,
$\conv(\{\lambda_i,\lambda_{i+2}\})$) and the line segment
$[\lambda_{i+1},\lambda_{i+3}]$. Each $a\in[a_i,a_{i+1}]$ also
lies in $[\lambda_{i+1},\lambda_{i+3}]$ and so has a unique
representation as a convex combination $\sum_j t_{ij}(a)\lambda_j$
with $t_{ij}(a)=0$ when $j\not\in\{i+1,i+3\}$. Likewise
$a\in[a_i,a_{i+1}]$ also lies in the eigentriangle $T(i,i+2,i+4)$
and so has a unique representation as a convex combination $\sum_j
s_{ij}(a)\lambda_j$ with $s_{ij}(a)=0$ when $j\in\{i+1,i+3\}$.

For each $a\in[a_i,a_{i+1}]$, let $p_i(a)=t_{i*}(a)-s_{i*}(a)$; as
in the proof of the last corollary, we see that $p_i(a)\in X$
where
$X=\{\vec1,\sigma,\overline{\sigma}\}^\perp\setminus\{\vec0\}$ and
that $a=f(p_i(a))$. Note that
$X\equiv\bbC^2\setminus\{\vec0\}\equiv\bbR^4\setminus\{\vec0\}$,
so that $X$ is simply connected. Moreover, each $p_i$ is
continuous on $[a_i,a_{i+1}]$. Because of the uniqueness of the
representations as convex combinations,
$t_{i*}(a_{i+1})=s_{(i+1)*}(a_{i+1})$ and
$s_{i*}(a_{i+1})=t_{(i+1)*}(a_{i+1})$. Thus
$p_i(a_{i+1})=-p_{i+1}(a_{i+1})$. Let
$\gamma_0:[0,10]\to\partial\Omega_2(U)$ be a (continuous) path
traversing $\partial\Omega_2(U)$ \emph{twice} in the
counterclockwise direction, beginning and ending at $a_1$ and such
that $\gamma_0([j,j+1])=[a_{j+1},a_{j+2}]$ ($j=0,1,\dots,9$) with
the understanding that the $a_i$ are numbered modulo 5 ($=N$). For
$t\in[j,j+1]$ let $\Gamma_0(t)=(-1)^jp_{j+1}(\gamma_0(t))$. The
alternating signs ensure that $\Gamma_0$ is continuous (even at
the integers), and the double circuit ensures that
$\Gamma_0(0)=\Gamma_0(10)$; in other words, that $\Gamma_0$ is a
\emph{loop} in $X$. Furthermore $f(\Gamma_0(t))=f(\pm
p_{j+1}(\gamma_0(t)))=\gamma_0(t)$.

Thus, given any $\lambda$ in the interior of $\Omega_2(U)$, the
winding number of $f\circ\Gamma_0$ relative to $\lambda$ is 2;
that is, $w_\lambda(f\circ\Gamma_0)=2$. Since $X$ is simply
connected, the loop $\Gamma_0$ is part of a continuous family of
loops $\Gamma_s$ ($0\leq s\leq1$) in $X$ such that $\Gamma_1$ is
the constant loop at some $p_*\in X$. Suppose $\lambda$ does not
lie on any of the loops $f\circ\Gamma_s$; then
$w_\lambda(f\circ\Gamma_s)=2$ for all $s$, a contradiction, since
$w_\lambda(f\circ\Gamma_1)=0$ ($f\circ\Gamma_1$ is the constant
loop at $f(p_*)$). Thus, for some $s,t$, $f(\Gamma_s(t))=\lambda$.
Since $\Gamma_s(t)\in X$, Lemma~\ref{sigmalemma} implies that
$\lambda\in\Sigma_2(U)$.
 \bx

This result has a number of consequences, as follows.

\begin{cor}\label{sigmacor1}
For every natural number $m$, Conj($5m,2m$) is valid.
\end{cor}

\Prf
Let $S_j=\{j,j+m,j+2m,j+3m,j+4m\}$ ($j=1,2,\dots,m$). The $S_j$
partition $\{1,2,\dots,5m\}$ into $m$ disjoint subsets. In view of
Lemma~\ref{omegaform}, we have
\[
   \Omega_{2m}(U)=\bigcap_{i=1}^{5m} D(i,i+2m,U)=
\]
\[
   \bigcap_{j=1}^m(\bigcap_{i\in S_j} D(i,i+2m,U))=
   \bigcap_{j=1}^m(\bigcap_{i=1}^5 D(i,i+2,U_j)),
\]
where $U_j$ is the unitary with spectrum
$\spec(U_j)=\{\lambda_i:i\in S_j\}$. Thus
$\Omega_{2m}(U)=\bigcap_{j=1}^m\Omega_2(U_j)$ and this is
$\bigcap_{j=1}^m\Lambda_2(U_j)$ by Theorem~\ref{52thm} (since
$|\spec(U_j)|=5$).

Consider any $\lambda\in\Omega_{2m}(U)$; for each $j=1,2,\dots,m$
this $\lambda\in\Lambda_2(U_j)$ and we have a 2--dimensional
subspace $\V_j$ of $\spn(\{\ket{\psi_i}:i\in S_j\})$ such that
$(U-\lambda I )\V_j\perp \V_j$. Now $\V_j$ and $(U-\lambda I)\V_j$
are subspaces of the mutually orthogonal
\[
   \spn\{\ket{\psi_i}:i\in S_j\}\,\,\,(j=1,2,\dots,m).
\]
Thus $\V=\V_1+\V_2+\dots+\V_m$ is a $2m$--dimensional subspace
such that $(U-\lambda I)\V \perp\V$, so that
$\lambda\in\Lambda_{2m}(U)$.
 \bx

\begin{rem} Although Corollary 6.4 ensures that Conj(15,6) is
correct, for instance, it may not be useful in applications
because it can happen that $\Omega_6(U)=\emptyset$ when $N=15$
($15=3\cdot6-3$; recall Theorem 4.6). Moreover, $\Omega_7(V)$ may
be empty in dimension 16 ($16\leq3\cdot7-3$) so that we cannot use
Proposition 4.5 to support the ``induction'' Conj(15,6)
$\Longrightarrow$ Conj(16,6). Thus Conj(16,6) remains undecided at
the moment.
\end{rem}

\begin{cor}\label{sigmacor2}
Conj($3k-1,k$) holds for all $k\geq 1$.
\end{cor}

\Prf The case $k=1$ is trivial (for all $N$). The case $k=2, N=5$
was proved in Theorem~\ref{52thm}. Based on this we can make an
induction on $k$ somewhat similar to that used in Theorem 5.2. For
$k+1>2$ consider $U$ with $|\spec(U)|=N=3(k+1)-1$. Let $W_1$
denote the wedge
\[
   W_1=D(1,1+(k+1),U)\cap D(1+(2k+1),1+(2k+1)+(k+1),U).
\]
Note that $\Omega_{k+1}(U)\subseteq W_1$ and in fact
$\Omega_{k+1}(U)$ is contained in the eigentriangle
$T_1=T(1,1+(k+1),1+(2k+1))$, since $(1+(2k+1))-(1+(k+1))=k<k+1$
and $1+(2k+1)+(k+1)=N+1\equiv1$. Let $U_1$ be the unitary with
spectrum
\[
   \spec(U_1)=\spec(U)\setminus\{\lambda_1,\lambda_{1+(k+1)},\lambda_{1+(2k+1)}\}.
\]
Then $\Omega_k(U_1)\supseteq \Omega_{k+1}(U))\cap D(1)$, where
$D(1)=D(1+k,1+(2k+2),U)$. Thus any $\lambda\in\Omega_{k+1}(U)\cap
D(1)$ is in $\Omega_k(U_1)$ as well as in $T_1$. Since
$|\spec(U_1)|=3k-1$ the inductive hypothesis ensures that
$\lambda\in\Lambda_k(U_1)$ and there is a $k$--dimensional
subspace $\V_1$ of $\spn(\{\ket{\psi_i}:\lambda_i\in\spec(U_1)\})$
such that $(U-\lambda I)\V_1\perp \V_1$. Since $\lambda\in T_1$ we
also have a 1--dimensional subspace $\V_1'$ of
$\spn\{\ket{\psi_1},\ket{\psi_{1+(k+1)}},\ket{\psi_{1+(2k+1)}}\}$
such that $(U-\lambda I)\V_1'\perp \V_1'$. The (orthogonal) sum
$\V_1+\V_1'$ shows that $\lambda\in\Lambda_{k+1}(U)$.

Similarly we have $D(k+3)=D(k+3+k,k+3+(2k+2),U)$ such that
$\lambda\in\Omega_{k+1}(U)\cap D(k+3)$ implies
$\lambda\in\Lambda_{k+1}(U)$. Finally, $D(1)\cup D(k+3)=\bbD$,
since $1+(2k+2)=k+3+k$ and $k+3+(2k+2)<N+1+k$ (if and only if $
k+1>2$).
 \bx

\begin{cor}\label{sigmacor3}
For $N=7$, we at least have
$\partial\Omega_3(U)\subseteq\Lambda_3(U)$.
\end{cor}

\Prf Along the lines of the proofs above, we need only show that
$\lambda\in\partial\Omega_3(U)$ implies
$\lambda\in\Omega_1(U')\cap\Omega_2(U'')$, where the spectra
$\sigma',\sigma''$ of $U',U''$ partition the spectrum $\sigma$ of
$U$, $|\sigma'|=2$, and $|\sigma''|=5$. Then
$\lambda\in\Lambda_1(U')$ and, using Theorem 6.3 we also have
$\lambda\in\Lambda_2(U'')$. To complete the argument note that, if
$\lambda$ belongs to one of the line segments forming a side of
$\partial\Omega_3(U)$, then $\lambda\in[\lambda_i,\lambda_{i+3}]$
for some $i$ and we set $\sigma'=\{\lambda_i,\lambda_{i+3}\}$.
Since $\lambda_{i+1}$ and $\lambda_{i+2}$ are the only points of
$\sigma''=\sigma\setminus\sigma'$ in the counterclockwise arc from
$\lambda_i$ to $\lambda_{i+3}$, we must also have
$\lambda\in\Omega_2(U'')$.
 \bx

\begin{rem} If there were some sort of \emph{a priori} convexity result
for the sets $\Lambda_k(T)$ -- along the lines of the
Hausdorff--Toeplitz Theorem for the classical numerical range
($=\Lambda_1(T)$) -- many of our arguments could be simplified and
extended. For example, from Corollary 6.6 we could derive
Conj(7,3). If Conj($N,k$) is true in general then we also have
convexity of all $\Lambda_k(T)$ for all normal $T$, but we do not
know of any independent argument for such convexity. In fact,
Conj(7,3) seems to be the ``smallest'' case that presently remains
unsettled.
\end{rem}

\section{Outlook}\label{S:conclusion}

As discussed above, the full Conjecture~A remains open. Here we
have constructively verified the conjecture in a wide variety of
cases, and, curiously, non-constructively in other cases. An
overarching conceptual proof covering all cases would be of great
interest. A possible avenue to such a result could come through a
general convexity theorem for higher-rank numerical ranges,
independent of normality or unitarity, though the present work
suggests that establishing such a result would be a delicate
matter. Progress in this direction is contained in the recent work
\cite{CGHK07}.

To apply the compression approach \cite{CKZ05b} to broader classes
of noise maps (in particular to randomized unitary channels with
more than two unitary errors) as a means to construct ideal
correctable codes, a better understanding is required of joint
solutions to the family of equations given by
Eqs.~(\ref{klcondition}). Furthermore, we have focussed on the
generic case of non-degenerate spectrum to streamline the
presentation. But many naturally arising physical examples include
degenerate spectra. There are extra technical issues to overcome
in such cases, but we expect our results can be extended to the
case of degenerate spectra.


It would be interesting to consider possible infinite-dimensional
extensions of these results. We have also not studied here
possible implications of the compression approach to more subtle
subsystem codes \cite{KLP05,KLPL05}. Nor have we considered
possible applications to approximate error-correction
\cite{BarKni02,CGS05,LNCY97,SchWes01}, and specifically to noise
maps that have two unitary errors which occur with high
probability.



{\noindent}{\it Acknowledgements.} M.D.C., J.A.H. and D.W.K. were
partially supported by NSERC. D.W.K. also acknowledges support
from ERA, CFI, and OIT. K. {\.Z}. acknowledges a partial support
by the grant number PBZ-MIN-008/P03/2003 of Polish Ministry of
Science and Information Technology.


%


\begin{thebibliography}{30}
\expandafter\ifx\csname
natexlab\endcsname\relax\def\natexlab#1{#1}\fi
\expandafter\ifx\csname bibnamefont\endcsname\relax
  \def\bibnamefont#1{#1}\fi
\expandafter\ifx\csname bibfnamefont\endcsname\relax
  \def\bibfnamefont#1{#1}\fi
\expandafter\ifx\csname citenamefont\endcsname\relax
  \def\citenamefont#1{#1}\fi
\expandafter\ifx\csname url\endcsname\relax
  \def\url#1{\texttt{#1}}\fi
\expandafter\ifx\csname
urlprefix\endcsname\relax\def\urlprefix{URL }\fi
\providecommand{\bibinfo}[2]{#2}
\providecommand{\eprint}[2][]{\url{#2}}















\bibitem{AL87}
\bibinfo{author}{\bibfnamefont{R.}~\bibnamefont{Alicki}}, \bibnamefont{and}
  \bibinfo{author}{\bibfnamefont{K.}~\bibnamefont{Lendi}},
  \textit{Quantum dynamical semigroups and applications},
  \bibinfo{journal}{Springer--Verlag, Berlin}, (\bibinfo{year}{1987}).

\bibitem{BarKni02}
\bibinfo{author}{\bibfnamefont{H.}~\bibnamefont{Barnum}},
  \bibinfo{author}{\bibfnamefont{E.}~\bibnamefont{Knill}},
   \bibinfo{journal}{{\it Reversing quantum dynamics with near-optimal quantum and classical fidelity},}
  \bibinfo{journal}{J. Math. Phys.} \textbf{\bibinfo{volume}{43}},
  \bibinfo{pages}{2097} (\bibinfo{year}{2002}).

\bibitem{BZ06} \bibinfo{author}{\bibfnamefont{I.}~\bibnamefont{Bengtsson}},
and
\bibinfo{author}{\bibfnamefont{K.}~\bibnamefont{{\.Z}yczkowski}},
\textit{Geometry of quantum states},
\bibinfo{journal}{Cambridge University Press} (\bibinfo{year}{2006}).



\bibitem{CGHK07}
\bibinfo{author}{\bibfnamefont{M.~D.}~\bibnamefont{Choi}},
  \bibinfo{author}{\bibfnamefont{M.}~\bibnamefont{Giesinger}},
  \bibinfo{author}{\bibfnamefont{J.~A.}~\bibnamefont{Holbrook}}, \bibnamefont{and}
  \bibinfo{author}{\bibfnamefont{D.~W.}~\bibnamefont{Kribs}},
   \textit{Geometry of higher-rank numerical ranges},
                \bibinfo{journal}{preprint, 2007.}

\bibitem{CKZ05b}
\bibinfo{author}{\bibfnamefont{M.~D.}~\bibnamefont{Choi}},
  \bibinfo{author}{\bibfnamefont{D.~W.}~\bibnamefont{Kribs}}, \bibnamefont{and}
  \bibinfo{author}{\bibfnamefont{K.}~\bibnamefont{{\.Z}yczkowski}},
     \bibinfo{journal}{{\it Quantum error correcting codes from the compression formalism},}
                \bibinfo{journal}{Rep. Math. Phys.}
                \textbf{\bibinfo{volume}{58}},
   \bibinfo{pages}{77-86} (\bibinfo{year}{2006}).

\bibitem{CKZ05a}
\bibinfo{author}{\bibfnamefont{M.~D.}~\bibnamefont{Choi}},
  \bibinfo{author}{\bibfnamefont{D.~W.}~\bibnamefont{Kribs}}, \bibnamefont{and}
  \bibinfo{author}{\bibfnamefont{K.}~\bibnamefont{{\.Z}yczkowski}},
\textit{Higher-rank numerical ranges and compression problems},
                \bibinfo{journal}{Lin. Alg. Appl.,}    \textbf{\bibinfo{volume}{418}},
   \bibinfo{pages}{828-839} (\bibinfo{year}{2006}).

\bibitem{CGS05}
\bibinfo{author}{\bibfnamefont{C.}~\bibnamefont{Crepeau}},
  \bibinfo{author}{\bibfnamefont{D.}~\bibnamefont{Gottesman}} \bibnamefont{and}
  \bibinfo{author}{\bibfnamefont{A.}~\bibnamefont{Smith}},
     \bibinfo{journal}{{\it Approximate quantum error-correcting codes and secret sharing schemes},}
  \bibinfo{journal}{quant-ph/0503139}.

\bibitem{Far93}
\bibinfo{author}{\bibfnamefont{D.~R.}~\bibnamefont{Farenick}},
   \bibinfo{journal}{{\it Matricial extensions of the numerical range: A brief survey},}
   \bibinfo{journal}{Linear and Multilinear Algebra}  \textbf{\bibinfo{volume}{34}},
   \bibinfo{pages}{197-211} (\bibinfo{year}{1993}).

\bibitem{Hal67}
\bibinfo{author}{\bibfnamefont{P.}~\bibnamefont{Halmos}},
\textit{A Hilbert space problem book},
   \bibinfo{journal}{D. Van Nostrand Company, Ltd., Toronto,} (\bibinfo{year}{1967}).

\bibitem{KL97a}
\bibinfo{author}{\bibfnamefont{E.}~\bibnamefont{Knill}} \bibnamefont{and}
  \bibinfo{author}{\bibfnamefont{R.}~\bibnamefont{Laflamme}},
     \bibinfo{journal}{{\it A theory of quantum error-correcting codes},}
  \bibinfo{journal}{Phys. Rev. {A}} \textbf{\bibinfo{volume}{55}},
  \bibinfo{pages}{900} (\bibinfo{year}{1997}).

\bibitem{KLP05}
\bibinfo{author}{\bibfnamefont{D.}~\bibnamefont{Kribs}},
  \bibinfo{author}{\bibfnamefont{R.}~\bibnamefont{Laflamme}} \bibnamefont{and}
  \bibinfo{author}{\bibfnamefont{D.}~\bibnamefont{Poulin}},
\textit{Unified and generalized approach to quantum error
correction},
  \bibinfo{journal}{Phys. Rev. Lett.} \textbf{\bibinfo{volume}{94}},
  \bibinfo{pages}{180501} (\bibinfo{year}{2005}).

\bibitem{KLPL05}
\bibinfo{author}{\bibfnamefont{D.~W.}~\bibnamefont{Kribs}},
  \bibinfo{author}{\bibfnamefont{R.}~\bibnamefont{Laflamme}},
  \bibinfo{author}{\bibfnamefont{D.}~\bibnamefont{Poulin}} \bibnamefont{and}
  \bibinfo{author}{\bibfnamefont{M.}~\bibnamefont{Lesosky}},
\textit{Operator quantum error correction},
 \bibinfo{journal}{Quantum Inf. \& Comp.} \textbf{\bibinfo{volume}{6}}
 \bibinfo{pages}{382} (\bibinfo{year}{2006}).

\bibitem{Lay}
\bibinfo{author}{\bibfnamefont{S.~R.}~\bibnamefont{Lay}},
\textit{Convex sets and their applications},
  \bibinfo{journal}{John Wiley \& Sons,} (\bibinfo{year}{1982}).

\bibitem{LNCY97}
\bibinfo{author}{\bibfnamefont{D.~W.}~\bibnamefont{Leung}},
  \bibinfo{author}{\bibfnamefont{M.~A.}~\bibnamefont{Nielsen}},
  \bibinfo{author}{\bibfnamefont{I.~L.}~\bibnamefont{Chuang}} \bibnamefont{and}
  \bibinfo{author}{\bibfnamefont{Y.}~\bibnamefont{Yamamoto}},
     \bibinfo{journal}{{\it Approximate quantum error correction can lead to better codes},}
 \bibinfo{journal}{Phys. Rev. A} \textbf{\bibinfo{volume}{56}}, \bibinfo{pages}{2567} (\bibinfo{year}{1997}).

\bibitem{LiTs91}
\bibinfo{author}{\bibfnamefont{C.-K.}~\bibnamefont{Li}}, \bibnamefont{and}
  \bibinfo{author}{\bibfnamefont{N.-K.}~\bibnamefont{Tsing}},
   \bibinfo{journal}{{\it On the $k$th matrix numerical range},}
   \bibinfo{journal}{Linear and Multilinear Algebra}  \textbf{\bibinfo{volume}{28}},
   \bibinfo{pages}{229-239} (\bibinfo{year}{1991}).

\bibitem{SchWes01}
\bibinfo{author}{\bibfnamefont{B.}~\bibnamefont{Schumacher}},
  \bibinfo{author}{\bibfnamefont{M.~D.}~\bibnamefont{Westmoreland}},
 \bibinfo{journal}{{\it Approximate quantum error correction},}
  \bibinfo{journal}{quant-ph/0112106}.







\end{thebibliography}
\end{document}